\documentclass[twocolumn,prb,aps,showpacs]{revtex4-1}
\usepackage{graphicx}
\usepackage{color}
\usepackage{dcolumn}
\usepackage{amsmath}
\usepackage{units}

\newcommand{\SIO}{Sr$_2$IrO$_4$ } 
\newcommand{\SIOns}{Sr$_2$IrO$_4$}

\begin{document}
\title{Observation of a $d$-wave gap in electron-doped \SIO}

\author{Y. K. Kim$^{1,\dagger}$, N. H. Sung$^{2,\dagger}$, J. D. Denlinger$^1$, B. J. Kim$^{2,*}$}
\address{$^1$Advanced Light Source, Lawrence Berkeley National Laboratory, Berkeley, CA 94720, USA}
\address{$^2$Max Planck Institute for Solid State Research, Heisenbergstra\ss e 1, D-70569 Stuttgart, Germany}

\maketitle

\noindent

\vspace{10 pt}
\vspace{10 pt}

{\bf High temperature superconductivity in cuprates emerges out of a highly enigmatic `pseudogap' metal phase. The mechanism of high temperature superconductivity is likely encrypted in the elusive relationship between the two phases, which  spectroscopically is manifested as Fermi arcs---disconnected segments of zero-energy states---collapsing into $d$-wave point nodes upon entering the superconducting phase.  Here, we reproduce this distinct cuprate phenomenology in the 5$d$ transition-metal oxide \SIOns. Using angle-resolved photoemission, we show that clean, low-temperature phase of 6-8$\%$ electron-doped \SIO has gapless excitations only at four isolated points in the Brillouin zone with a predominant $d$-wave symmetry of the gap. Our work thus establishes a connection between the low-temperature $d$-wave instability and the previously reported high-temperature Fermi arcs in electron-doped \SIO (ref.~1). Although the physical origin of the $d$-wave gap remains to be understood, \SIO is a first non-cuprate material to spectroscopically reproduce the complete phenomenology of the cuprates, thus offering a new material platform to investigate the relationship between the pseudogap and the $d$-wave gap.}

The one-to-one correspondence between \SIO and high-temperature supreconducting (HTSC) cuprates in their electronic\cite{Kim08}, magnetic\cite{Kim09,Kim12,Fujiyama12,Jackeli09}, and lattice structures\cite{Crawford} allows us to present in Fig.~1a and 1b the angle-resolved photoemission (ARPES) intensity maps of $\approx$7$\%$ electron-doped \SIO recorded at the Fermi level following the standard notations used in the cuprate literature. A sign difference between the two systems in one of the tight-binding parameters\cite{Jin} characterizing the hopping of an electron in a quasi-two-dimensional (2D) square lattice renders Fermi surfaces shifted by ($\pi$,$\pi$) with respect to each other in their non-interacting electron descriptions. This sign of the next-nearest hopping can be reversed by electron-hole conjugation, which means that our results on electron-doped \SIO can be directly compared to those of hole-doped cuprates\cite{Wang}. 

At T=10 K, the map shows a locus of finite intensities, which is largely concentrated at a point near ($\pi$/2,$\pi$/2). Upon inspection of the energy distribution curves (EDC) along this locus, plotted in Fig.~1c after symmetrization with respect to the Fermi level following standard procedures\cite{Norman}, it is clearly seen that the spectra are gapped everywhere except at a single momentum ($k$) point in a quadrant of the Brillouin-zone. This `nodal' Fermi surface is completely unexpected from the single-electron band structure, which implies that a highly non-trivial, correlated electronic phase develops upon electron doping of \SIOns.

\begin{figure}
\centerline{\includegraphics[width=1\columnwidth,angle=0]{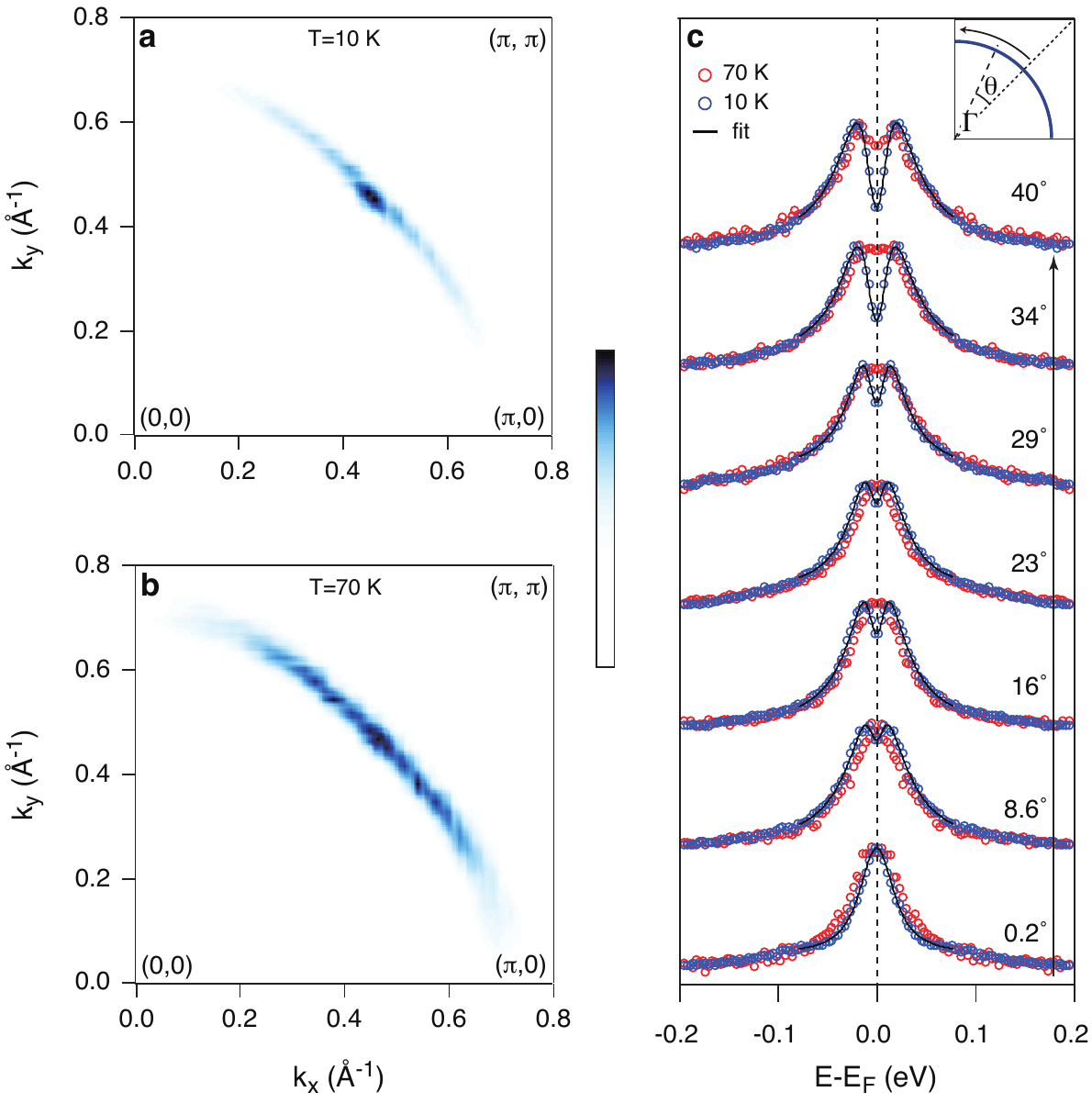}}
\caption{{\bf Low-temperature nodal Fermi surface and high-temperature Fermi arcs.} ARPES intensity maps at the Fermi level for $\approx$7$\%$ electron-doped \SIO measured at (a) T=10 K and (b) T=70 K. The images were symmetrized with respect to the (0,0)-($\pi$,$\pi$) line. (c) Symmetrized EDCs (open circles) along the underlying Fermi surface. T=10 K data (blue open circles) were fitted (solid lines) to the dynes Formula. Inset shows the definition of the angle $\theta$ that labels the EDCs. $\approx$7$\%$ electron doping level was reached using in situ doping method by deposing 0.8 monolayer (ML) of potassium atoms on a freshly cleaved sample surface\cite{Kim14}.}\label{fig:fig1}
\vspace*{-0.5cm}
\end{figure}

\begin{figure*}
\hspace*{-0.2cm}\vspace*{-0.1cm}\centerline{\includegraphics[width=1.7\columnwidth,angle=0]{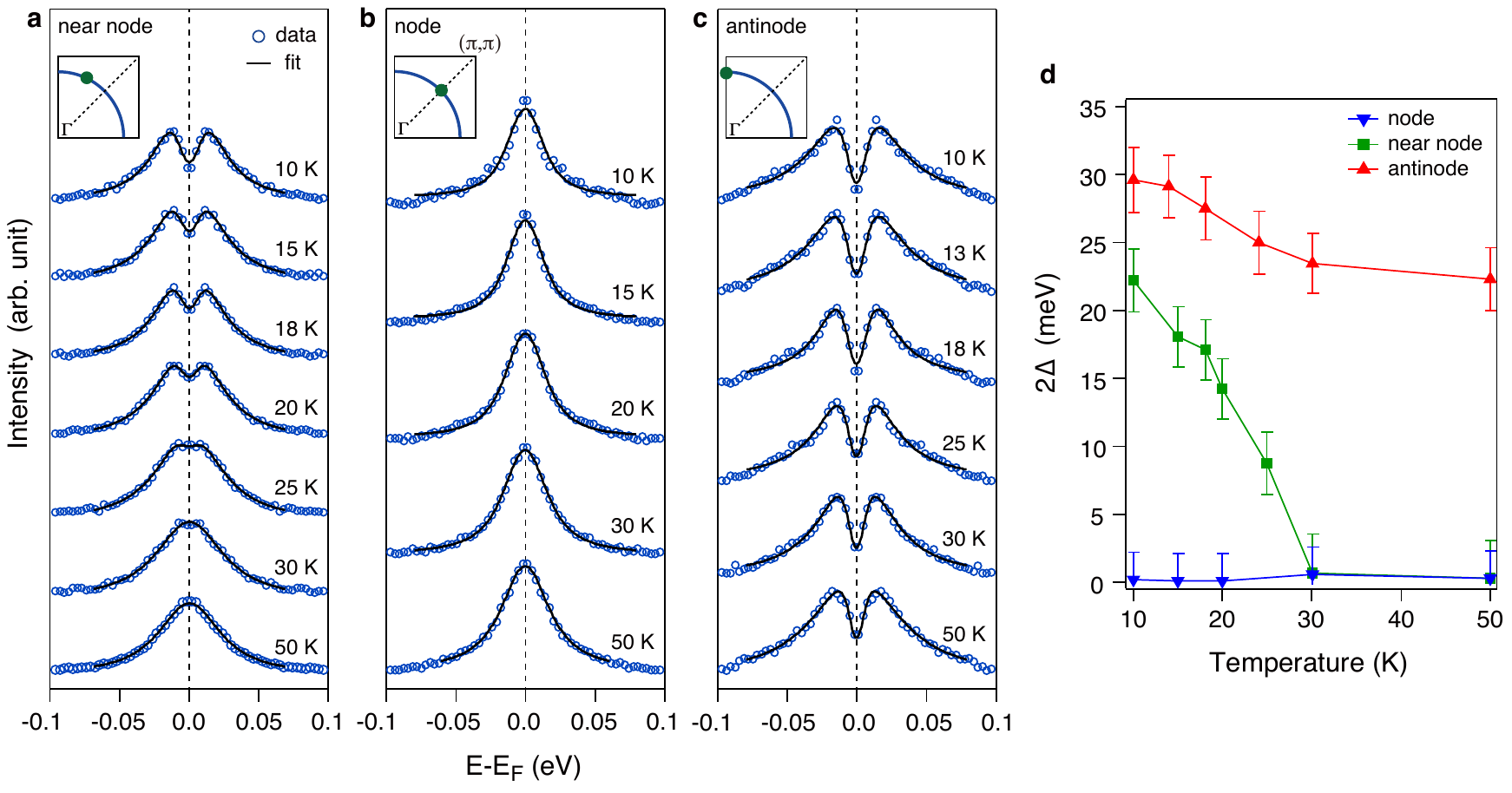}}

%
\caption{{\bf Temperature dependence of the gap.} The data were collected for three representative momenta. Symmetrized EDCs (open circles) are fit to the Dynes formula (solid lines) for (a) near-node spectra (17$^\circ$ from the node),  (b) nodal spectra, and (c) antinodal spectra. (d) The gap magnitude as a function of temperature. Error bars represent standard error in the fitting procedure. 
}\label{fig:fig2}
\vspace*{-0.5cm}
\end{figure*}

Density-functional calculations of \SIO predict a circular electron Fermi surface centered at the $\Gamma$ point whose enclosed area is half of the 2D Brillouin zone\cite{Kim08}. Indeed, such a `large' Fermi surface is observed at a sufficiently high doping and/or temperature\cite{Kim14}, implying that the nodal Fermi surface arises in close proximity to the Mott insulating, ($\pi$,$\pi$) antiferromagetic ground state of the parent \SIO (ref.~3). Thus, the previously reported high-temperature Fermi arcs interpolate between these two contrasting electronic phases:  the node evolves to a Fermi arc of a finite length at high temperature (Fig.~1b) as the gap closes along this arc (Fig.~1c). This establishes a non-trivial origin the Fermi arcs, which was indicated by the strong temperature and doping dependence of the arc length, and its position displaced from the large Fermi surface\cite{Kim14}. These features distinguish the Fermi arcs in electron-doped \SIO from other Fermi-arc-like features in manganites\cite{Mannella} or nickelates\cite{Uchida}, which merely show anisotropic suppression of spectral weights at the Fermi level but otherwise are consistent with their density-functional descriptions. 

\begin{figure*}
\hspace*{-0.2cm}\vspace*{-0.1cm}\centerline{\includegraphics[width=1.4\columnwidth,angle=0]{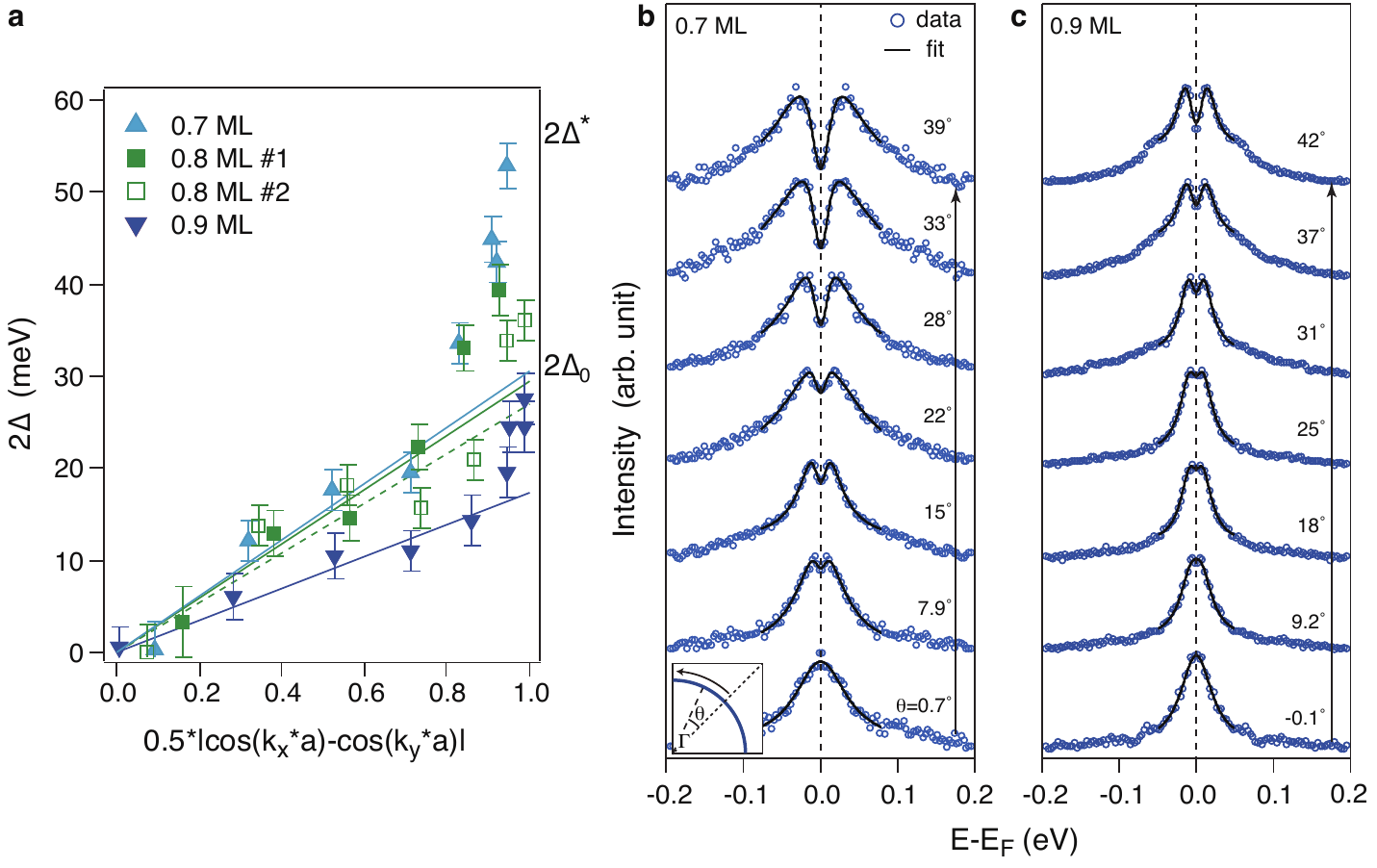}}

\caption{{\bf Momentum dependence of the gap at T=10 K.}  The data were collected on four samples of varying electron doping level (potassium coverage). (a) A compilation of the extracted gaps as a function of $d$-wave form factor. Solid and dashed lines indicate linear fits to the gap magnitude in the near-nodal region. Two samples for 0.8 ML coverage gave consistent results. Error bars represent standard error in the fitting procedure. The gap magnitude was extracted by fitting the symmetrized EDC (circles) to the Dynes formula (solid lines). Symmetrized EDCs are shown for (b ) 0.7 ML and (c) 0.9 ML, which approximately are estimated to be in the doping range of 6-8$\%$ based on the previous study\cite{Kim14}. Inset in (b) shows the definition of the angle that labels the EDCs. 
}
\vspace*{-0.5cm}
\end{figure*}

To further investigate the connection between the Fermi arcs and the nodal Fermi surface, we follow in Fig.~2a the temperature evolution at a $k$ point on the Fermi arc, 17$^\circ$ away from the node. With decreasing temperature, a finite gap (2$\Delta$) is resolved below T=30 K reaching a maximum of $\approx$22 meV at T=10 K (Fig.~2d). The temperature evolution of the spectra at the node (Fig.~2b) and the antinode (Fig.~2c) is insensitive to the near-nodal gap opening, indicating that the near-nodal gap is a defining feature that differentiates between the low-temperature nodal metal phase and the high-temperature Fermi arc phase. This $k$-dependent temperature evolution of the spectra is strikingly reminiscent of that observed in the cuprates\cite{Norman,Lee}.

 It is unclear based on our data whether or not the gap opening is associated with a phase transition.  In superconducting cuprates, the near-nodal gap follows a canonical temperature dependence of a Bardeen-Cooper-Schrieffer order parameter\cite{Lee}. In non-superconducting cuprates, such as the stripe compound La$_{2-x}$Ba$_x$CuO$_4$ with one-eighth doping\cite{Valla} and a highly-underdoped Bi$_2$Sr$_2$CaCu$_2$O$_{8+\delta}$\cite{Chatterjee}, the Fermi arcs continuously shrink to $d$-wave nodes without encountering a phase transition.  In either case, nodal fermions arise from $d$-wave pairing correlations, which regardless of presence of phase-coherent superconductivity results in a $d$-wave gap in the single-particle excitation spectra\cite{Balents}. However, it is also plausible that some other phase distinct from superconductivity leads to a similar gap structure. For example, $d$-density wave, predicted long ago to compete with $d$-wave superconductivity, may be consistent with our data\cite{Chakravarty, Affleck}.

Although we are not able to discriminate among such possible scenarios, the symmetry of the gap structure places a stringent constraint on the possible underlying electronic phases. Figure 3a documents the gap extracted from fitting to the Dynes formula as a function of the $d$-wave form factor $|$$\cos$($k_x$)$\--$$\cos$($k_y$)$|$/2. We have repeated the measurements on two samples with same nominal doping of $\approx$7$\%$, and two other samples with slightly lower (Fig.~3b) and higher dopings lower (Fig.~3c). In all four measured samples, we find a good agreement with the $d$-wave form factor near the node, but a prominent deviation from it in the anti-nodal region. A linear extrapolation of the near-node gaps returns a $d$-wave gap maximum in the range 16$\leq$ $2\Delta_0$ $\leq$30 meV, which is smaller roughly by a factor of two compared to the actual gaps $2\Delta^*$ measured at the antinode. 

We note that this particular gap profile deviating from a simple $d$-wave form is observed across many different families of underdoped cuprates, suggesting that the near-nodal gap and the anti-nodal gap have two different origins\cite{Hashimoto, Yoshida}. This in turn connects to the elusive relationship between the superconducting and the pseudogap phase, which is one of the outstanding issues in the field of HTSC widely believed to be important for unraveling its mechanism. Although the nature of the $d$-wave instability in electron-doped \SIO remains to be understood, the striking parallel between the two microscopically disparate systems points to a common origin of the pseudogap that arises in close proximity to a $d$-wave instability.

Some remark is in order regarding the origin of the $d$-wave gap in \SIOns. We believe that it is most likely associated with HTSC. In fact, to the best of our knowledge, all observations of $d$-wave gaps to date involve pairing of electrons. $d$-wave superconductivity is also expected from theoretical studies\cite{Wang,Watanabe,Meng} of electron-doped \SIOns. On the other hand, it is well known that doping a spin-1/2 antiferromagnetic Mott insulator generates a plethora of electronic orders\cite{Keimer,Fradkin}, such as spin/charge density waves, electron liquid crystals, and loop current orders\cite{Varma,Bourges}; and therefore it cannot be ruled out that the $d$-wave gap in \SIO represents an entirely new quantum state of matter that competes with $d$-wave superconductivity. For guidance to future identification of the precise nature of the $d$-wave nodal metal, we note that well-defined quasiparticles, sharper at higher doing, are observed along the entire $k$-trajectory from the node to the antinode, which implies that the $d$-wave nodal metal phase supports a coherent charge excitation without a significant momentum anisotropy. We note that the antinode quasiparticle intensity in cuprates has been shown to correlate with the superfluid density\cite{Feng}. 

We now turn to describing our unique experimental approach that allowed ARPES measurements of clean, low-temperature phase of electron-doped \SIOns. Unlike cuprates, chemical doping of \SIO turns out to be challenging; we are not aware of any ARPES data measured on a chemically doped sample\cite{YCao,Fisher,Baumberger} that show sharp quasiparticles, which is a measure of a coherent charge transport, let alone a $d$-wave gap. The in situ doping method provides an alternative to conventional chemical doping, and has been successfully applied across many different systems\cite{Ohta, Hossain, Zhang} including HTSC cuprates\cite{Hossain}. 

This method involves evaporating alkali metals on a clean sample surface prepared in a ultra-high vacuum, and induces electron doping without disrupting the host material, via diffusion of alkali valence electrons into the sample surface. In an earlier study using a combination of in situ doping and ARPES, the single-particle spectral function for the entire doping range from the Mott insulator to a Fermi liquid metal on the electron-doped side has been revealed using a parent insulator \SIOns\cite{Kim14}. However, a drawback of this method is that doping is limited to few surface layers and the bulk of the sample is left undoped so that standard bulk-sensitive probes are not applicable. Moreover, because insulators are intrinsically difficult to measure using single-particle probes due to sample charging, accessible temperature range is limited as the sample becomes more insulating at lower temperatures. This limits the ARPES measurement temperature above T$\approx$70 K for pristine \SIOns, leaving the low-temperature region of the phase diagram unrevealed.  

\begin{figure}
\centerline{\includegraphics[width=1\columnwidth,angle=0]{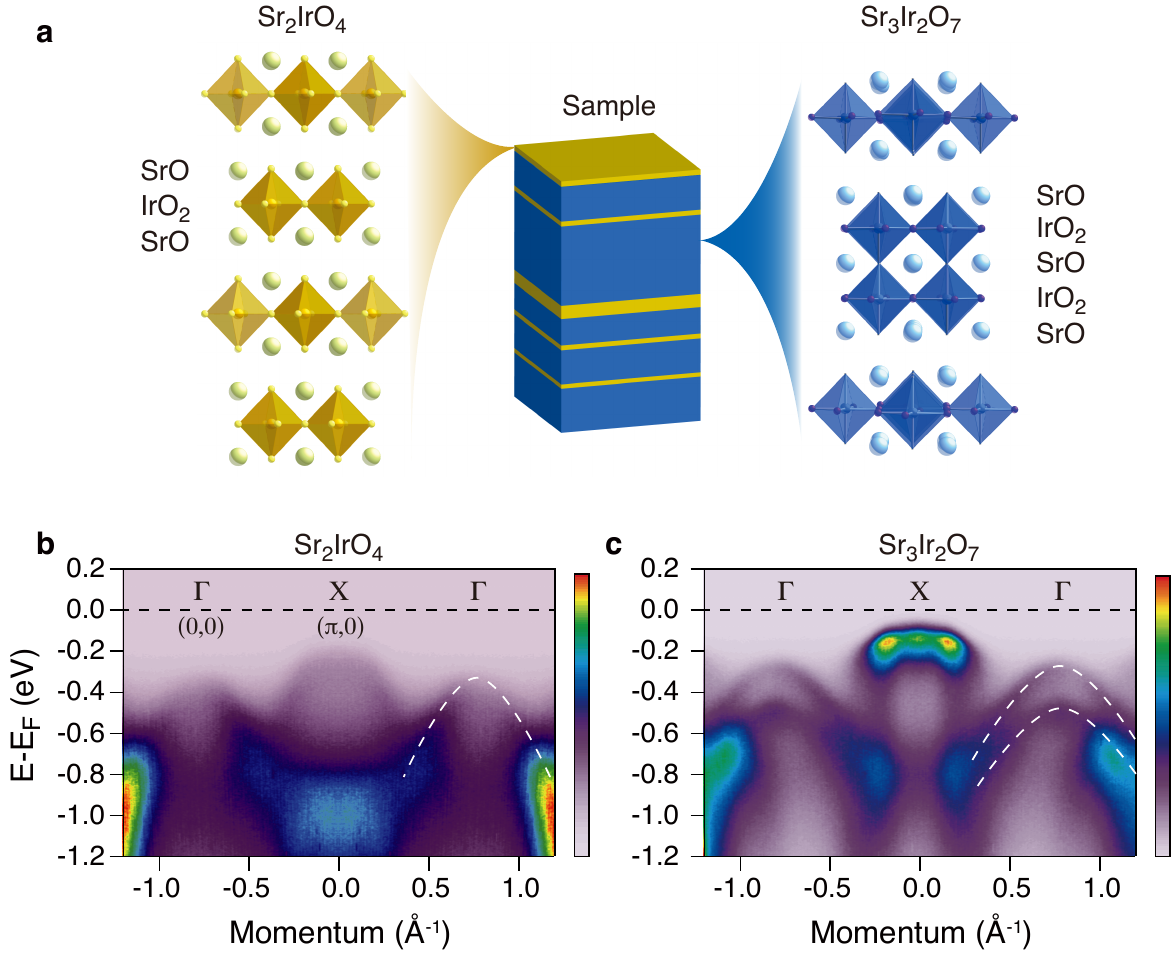}}
\caption{{\bf Illustration of our materials scheme to access low-temperature spectral functions for electron-doped \SIOns.} (a) Few-layers-thick slabs of \SIO are intergrown into the double-layer member of the Ruddlesden-Popper series Sr$_{n+1}$Ir$_n$O$_{3n+1}$, Sr$_3$Ir$_2$O$_7$, which has resistivity at T=10 K many orders of magnitude lower\cite{Cao} than that in the single layer \SIOns\cite{Kim09} and is structurally compatible with \SIOns. \SIO and Sr$_3$Ir$_2$O$_7$ can be intergrown into each other to an arbitrary ratio of single-layer to double-layer phase. Our crystals had 20-30$\%$ number fraction of iridium (IV) ions in the single-layer phase (Fig.~S1), optimized to maximize the probability of exposing the single-layer phase when cleaved while avoiding sample charging. The single-layer phase exposed at the topmost surface layer after cleaving can easily be identified through the absence of a bilayer splitting (dashed lines) along the $\Gamma$-X line spectra, shown in (b)  and (c), for \SIO and Sr$_3$Ir$_2$O$_7$, respectively.}\label{fig:fig1}
\vspace*{-0.6cm}
\end{figure}

To circumvent the charging problem, we inserted few-layers-thick slabs of \SIO into its more conducting sister compound Sr$_3$Ir$_2$O$_7$ (ref.~33) as illustrated in Fig.~4. We did not find any sign of change in the physical properties due to interfacing the two members of the Ruddlesden-Popper series (Fig.~S1). Indeed, we were able to reproduce the previously reported high-temperature Fermi arcs in our `engineered' sample (Fig.~S2). As demonstrated in Fig.~2, our approach allows ARPES measurement down to the base temperature of T=10 K. At the node (Fig.~2b), absence of a gap in the EDC confirms that there is no charging down to T=10 K. 

Our work shows that when clean electron doping is achieved, \SIO displays a correlated electron phenomenon expected for a doped (pseudo)spin-1/2 antiferromagnetic Mott insulator, whose material realization has thus far been limited to cuprates. In order for the nature of the $d$-wave nodal metal to be scrutinized using the full arsenal of condensed matter physics, the observations presented herein must be reproduced in a chemically doped sample. This will also directly verify, or disprove, the superconducting origin of the $d$-wave gap in electron-doped \SIOns. Either way, our work represents a discovery of a novel quantum state in the emerging family of 5$d$ magnetic insulators. With their diverse spectrum of magnetism beyond the conventional Heisenberg paradigm\cite{bilayer,honeycomb}, they open a new route to investigate the elusive connection between the spin and charge degrees of freedom, long believed to lead to high temperature superconductivity.

\noindent
 We acknowledge helpful discussions with C. Kim, G. Khaliullin, B. Keimer, M. Le Tacon, G. Jackeli, J. F. Mitchell, M. Norman and J. W. Allen. We thank B. Y. Kim for technical assistance. The Advanced Light Source is supported by the Director, Office of Science, Office of Basic Energy Sciences, of the U.S. Department of Energy under Contract No. DE-AC02-05CH11231. Y. K. Kim is supported through NRF Grants funded by the MEST (No. 20100018092).

\noindent
$^\dagger$These authors contributed equally to the project.
$^*$Correspondence and requests for materials should be addressed to  B.J.K.~(bjkim@fkf.mpg.de).





\vspace{10 pt}

\noindent\textbf {Supplementary Information}\\\\
\vspace{10 pt}
\noindent
{\bf A. Angle-resolved photoemission spectroscopy}

\noindent
ARPES measurements were performed at Beamline 4.0.3 at the Advanced Light Source, equipped with a Scienta R8000 electron analyzer. We evaporated potassium atoms in situ onto the freshly cleaved surface of the sample using a commercial SAES evaporator to dope electrons into the sample surface. The samples were cleaved and doped at T=70 K and p$\sim$ 3$\times$10$^{-11}$ Torr, and measured in the temperature range T=10-70 K with an overall energy resolution of 14 meV at $h\nu$=68 eV. Because of the short lifetime of the sample after in situ doping ($\sim$1 hour), data shown in Figs.~1-3 were collected on 7 different samples. The reproducibility of the surface coverage level was about $\pm$0.05 ML. The surface coverage of potassium (K) was monitored through the K 3p core level and the K overlayer quantum well band as described in Ref.~4. 

\begin{figure}[b]
\hspace*{-0.2cm}\vspace*{-0.1cm}\centerline{\includegraphics[width=0.7\columnwidth,angle=0]{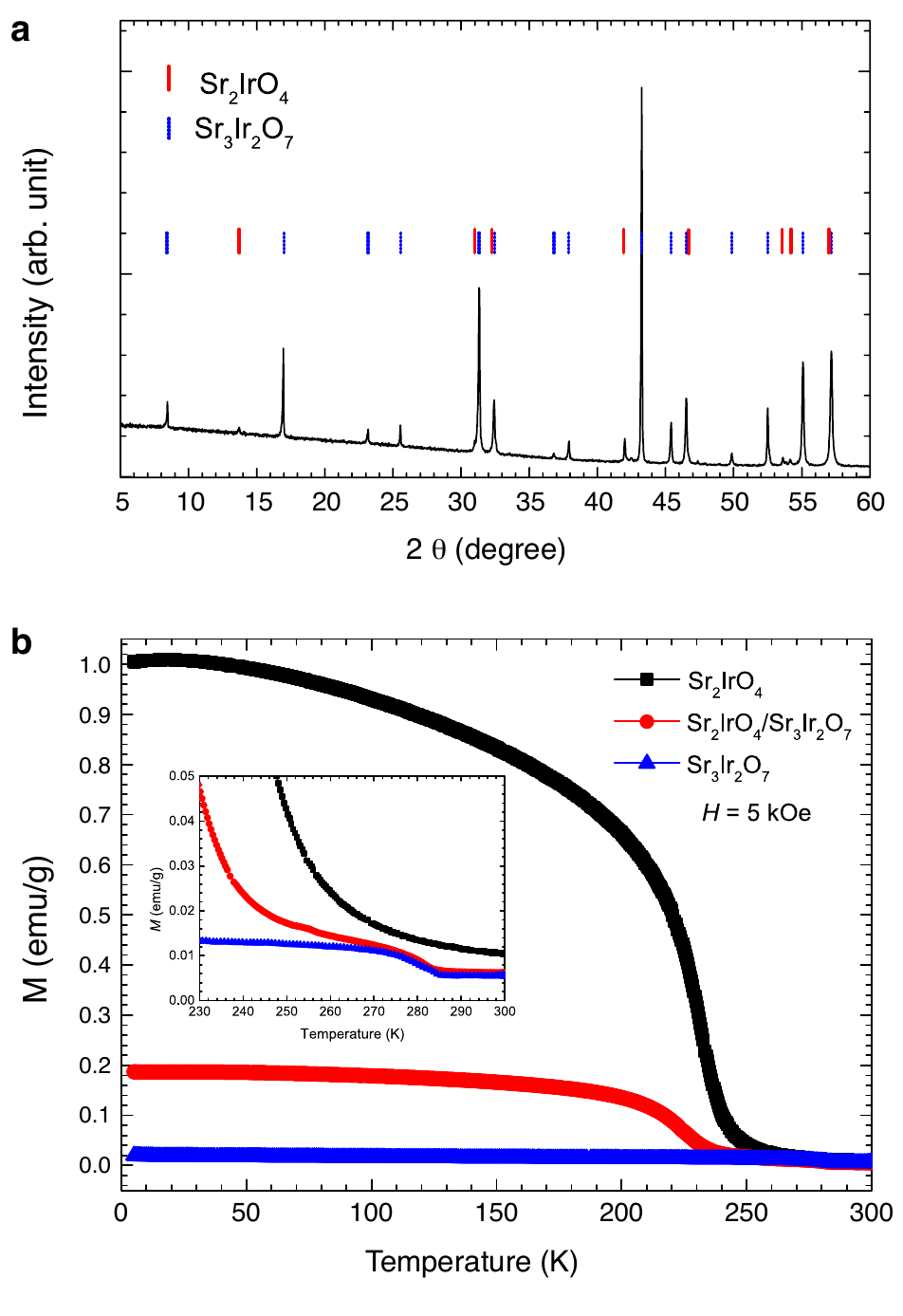}}
\setcounter{figure}{0}
 \renewcommand{\thefigure}{S\arabic{figure}}

\caption{{\bf Sample characterization.} (a) Powder x-ray diffraction pattern collected on ground crystals. (b) Temperature dependence of the magnetization. Data were recorded on warming after field cooling at 5 kOe applied in the ab-plane direction. Inset shows a zoom into the T$_{\textrm N}$$\approx$285 K of the `host' crystal Sr$_3$Ir$_2$O$_7$. 
}\label{fig:fig3} 
\end{figure}

\vspace{10 pt}
\noindent
{\bf B. Sample growth and characterizations}

\noindent
The samples were grown by the flux method. Sr$_2$CO$_3$, IrO$_2$, and SrCl$_2$ were mixed in a molar ratio of 2:1:7, melted and soaked at T=1125$^\circ$C, and then slow-cooled down to 880$^\circ$C. The degree of inter-growth of \SIO phase was controlled by varying the soaking temperature in the range 1050$^\circ$C-1300 $^\circ$C. The powder x-ray diffraction pattern (Fig.~S1a) showed a two-phase mixture of \SIO and Sr$_3$Ir$_2$O$_7$ and no other phases. The ratio between the two phases could readily be read off from the value of the saturation magnetization measured by SQUID magnetometry (Fig.~S1b) since Sr$_3$Ir$_2$O$_7$ phase a negligible magnetization as compared to \SIOns. A pure \SIO phase has a saturation moment of approximately 1 emu/g. The magnetic ordering temperatures for both \SIO (240 K) and Sr$_3$Ir$_2$O$_7$ (285 K) phases in the sample were unchanged by interfacing the two phases.

\begin{figure}
\hspace*{-0.2cm}\vspace*{-0.1cm}\centerline{\includegraphics[width=0.7\columnwidth,angle=0]{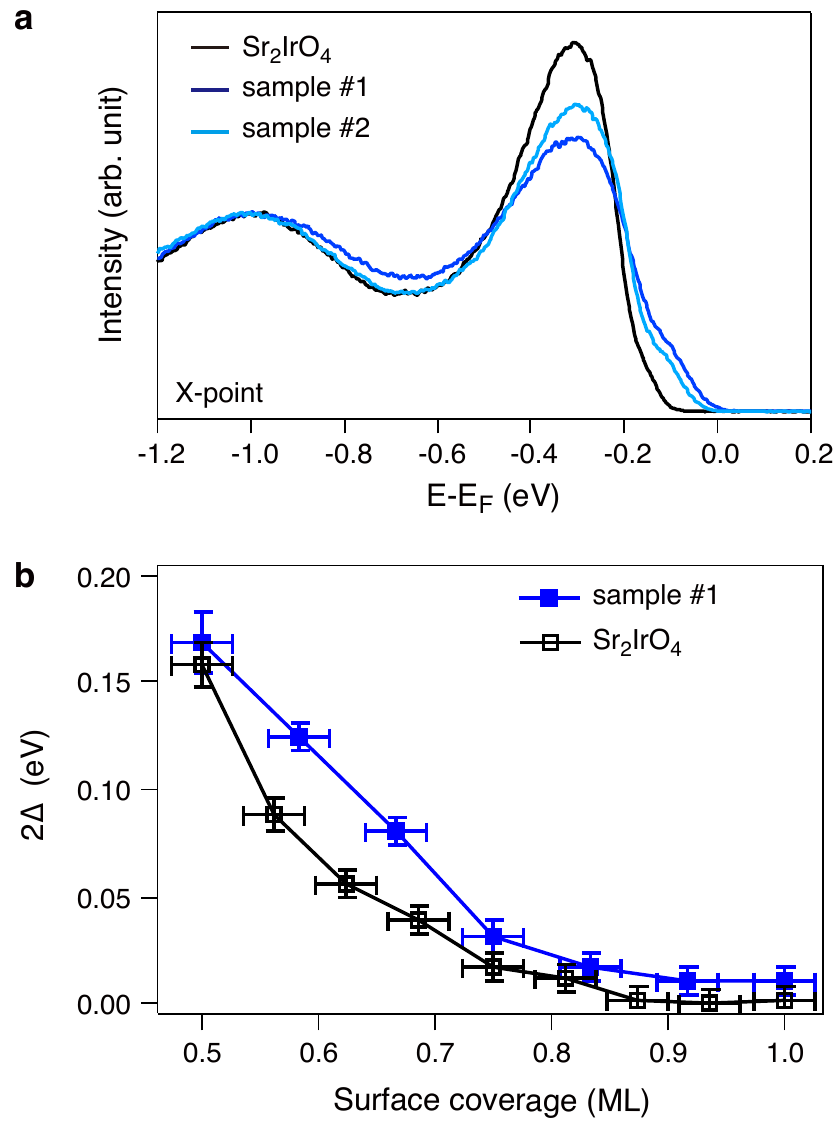}}
 \renewcommand{\thefigure}{S\arabic{figure}}
\caption{{\bf Comparison to the pristine \SIOns.} (a) Angle-resolved spectrum at the X point.  (b) Surface coverage dependence of the pseudogap at the antinode at T=70 K.
}
\end{figure}

The slabs of \SIO phase were thin enough to allow the Sr$_3$Ir$_2$O$_7$ lying underneath to be seen in the ARPES spectrum, which typically has a probing depth of about $\lesssim$ 1 unit cell. A small `foot' is seen at $\approx$0.1 eV at the X point (Fig.~S2a), which is from the Sr$_3$Ir$_2$O$_7$ phase having a smaller binding energy of the valence band at the X point (see Fig.~4b and 1c). The pseudogap behaviour reported earlier was well reproduced in our sample (Fig.~S2b). Subtle differences, though, are observed; the gap tends to be slightly larger and stays open all the way up to 1 ML. We note that this could be due to a difference in the depth profile of doped carriers.

\end{document}